\documentclass[aps,pra,twocolumn,10pt,groupedaddress,nofootinbib,superscriptaddress,notitlepage,showpacs,floatfix]{revtex4-1}
\usepackage{graphicx,graphics}
\usepackage{subfigure}
\usepackage{dcolumn}   
\usepackage{array}
\usepackage{amsthm}
\usepackage{amsmath}
\usepackage{amssymb}
\usepackage{amsfonts}
\usepackage{mathrsfs} 
\usepackage{bm}
\usepackage{braket} 
\usepackage{enumitem}
\usepackage{url}
\usepackage[pdfstartview=FitH,pdfencoding=auto]{hyperref}
\usepackage{bookmark}
\usepackage{pifont}
\hypersetup{
    colorlinks=true,        		
    linkcolor=blue,              	
    citecolor=blue,        			
    filecolor=blue,          		
    urlcolor=blue,              	
    runcolor=cyan
			}
\usepackage[pdftex]{color}
\usepackage{multirow}
\usepackage{CJK}
\usepackage{cancel}
\usepackage{ulem}

\begin{document}
\begin{CJK*}{GB}{}
	\title{Beating standard quantum limit via 
		two-axis magnetic susceptibility measurement}
\author{Zheng-An Wang}
	\thanks{The two authors contributed equally to this work.}
\affiliation{Institute of Physics, Chinese Academy of Sciences, Beijing 100190, China}
\affiliation{School of Physical Sciences, University of Chinese Academy of Sciences, Beijing 100190, China}
\author{Yi Peng}
	\thanks{The two authors contributed equally to this work.}
\affiliation{Institute of Physics, Chinese Academy of Sciences, Beijing 100190, China}
\affiliation{ Shenzhen Insititute for Quantum Science and Engineering,
Southern University of Science and Technology, Shenzhen 518055, China}
\author{Dapeng Yu}
\affiliation{ Shenzhen Insititute for Quantum Science and Engineering,
Southern University of Science and Technology, Shenzhen 518055, China}
\affiliation{Guangdong Provincial Key Laboratory of Quantum Science and Engineering, Southern University of Science and Technology, Shenzhen, 518055, China}
\affiliation{Shenzhen Key Laboratory of Quantum Science and Engineering, Southern University of Science and Technology, Shenzhen,518055, China}
\author{Heng Fan}
\email{hfan@iphy.ac.cn}
\affiliation{Institute of Physics, Chinese Academy of Sciences, Beijing 100190, China}
\affiliation{School of Physical Sciences, University of Chinese Academy of Sciences, Beijing 100190, China}
\affiliation{CAS Center for Excellence in Topological Quantum Computation, University of Chinese Academy of Sciences, Beijing 100190, China}
\affiliation{Songshan Lake Materials Laboratory, Dongguan 523808, Guangdong, China}
\date{\today}
\begin{abstract}
	We report a metrology scheme which measures magnetic susceptibility of an atomic 
    spin ensemble along the $x$ and $z$ direction and produces parameter estimation with
    precision beating the standard quantum limit. The atomic ensemble is initialized via
    one-axis spin squeezing with optimized squeezing time and parameter $\phi$ to be 
    estimated is assumed as uniformly distributed between 0 and $2\pi$. One estimation 
    of $\phi$ can be produced with every two magnetic susceptibility data measured along the two axis respectively, 
    which has imprecision 
    scaling $({1.43\pm0.02})/N^{0.687\pm0.003}$ with respect to the number $N$ of 
    atomic spins. The measurement scheme is easy to implement and thus one step towards
    practical application of quantum metrology.
\end{abstract}
\maketitle
\end{CJK*}
\section{Introduction.}
Metrology is the cornerstone of scientific research and technology development. Quantum 
mechanics provides new opportunities to metrology and attracts attentions from different 
fields~\cite{giovannetti2001quantumenhanceda,jozsa2000quantuma,taylor2013biological,
schnabel2010quantum,ligo2013enhanced}. Given $N$ probes, the lowest parameter estimation
imprecision allowed by classical theory is of the order of $1/\sqrt{N}$, which is the 
so-called standard quantum limit (SQL). It comes from the well-known central limit 
theorem. Quantum metrology harnesses quantum features such quantum entanglement and spin
squeezing to beat the standard quantum limit. The lowest parameter estimation 
imprecision can be of the order $1/N$ when quantum effects are fully utilized and thus 
result in a quadrature enhancement of estimation precision. It is the so-called 
Heisenberg limit (HL)~\cite{braunstein1994statistical,giovannetti2006quantum}. 

Every metrology scheme involves three stages: 1) prob state initialization; 2) sensing 
parameter field; 3) measurement of probes and information extraction. Among them, the 
second stage depends on the parameter field to be estimated. What we can do is to 
choose suitable prob systems and design initialization and measurement procedure 
carefully, which correspond to the first stage and the final stage design respectively. 
Quantum entanglement and spin squeezing are generated in atomic ensembles which can host
$10^3\sim10^5$ atoms. It is a platform of great potential in quantum metrology, because 
of its sheer size as well as the strong quantum entanglement that can be generated 
between so many probes with technology available to many 
labs~\cite{riedel2010atomchipbased,gross2010nonlinear,lucke2011twin,strobel2014fisher}.
Atoms in ensembles are indistinguishable. Thus neither manipulation nor measurement 
on individual probe is possible as required in many proposed metrology 
schemes~\cite{berry2000optimal,hentschel2010machineb,hentschel2011efficient,lovett2013differential,peng2020feedback}.
Magnetic susceptibility $\hat{J}_n$ along a specific axis is the most common and 
convenient physical quantity to be measured of an atomic ensemble. When the average 
magnetic susceptibility $\braket{\hat{J}_n}$ is used to estimate the parameter field
$\phi$, the corresponding estimation imprecision would be 
$\Delta\phi
={\Delta{}J_n}\Big/\left|\frac{\partial\braket{\hat{J}_n}}{\partial{}\phi}\right|$. 
Here $\Delta{}J_n=\sqrt{\braket{(\hat{J}_n-\braket{\hat{J}_n})^2}}$ is the standard 
deviation of the magnetic susceptibility along direction $n$.
Thus great fluctuation $\Delta{}J_n$ would result in poor estimation. It is widely 
believed that ``low-noise detection" is necessary to improve 
parameter estimation 
precision~\cite{riedel2010atomchipbased,gross2010nonlinear,lucke2011twin,strobel2014fisher}.
\begin{figure*}[!htb]
	\includegraphics[width=\textwidth]{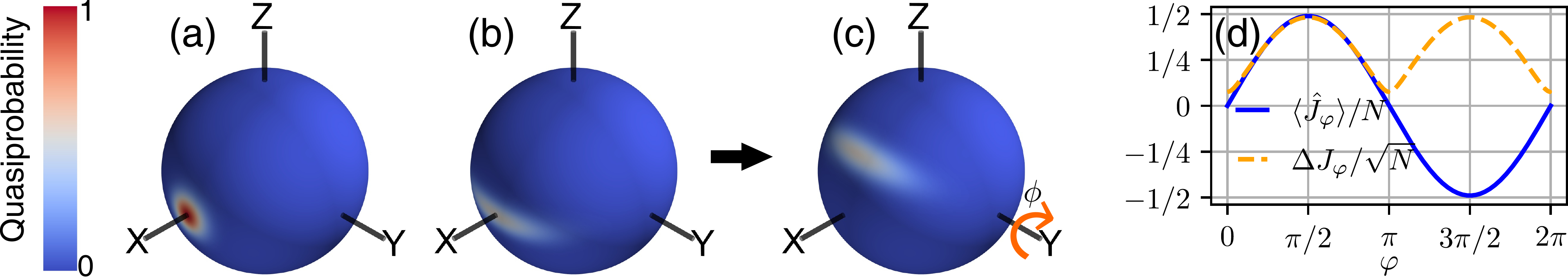}
    \caption{(Color online) Quasiprobability and magnetic susceptibility. The quasiprobability $P(\theta,\phi)=\left|\braket{\theta,\phi|\psi}\right|^2$ has been calculated for spin 
	ensemble with $N=100$ spins. $\ket{\theta,\phi}$ is the coherent spin state where all the spins pointing along the direction with 
	polar angle $\theta$ and azimuth angle $\phi$. 
	The distribution of the quasiprobability is of similar shape for other values of $N$.
	The only difference is that the quasiprobability would be more concentrated as the number of spins grows 
	bigger. (a) The coherent spin state is initialized along the positive $x$ direction. (c) is the one-axis
	spin squeezed state as shown in (b) rotated along the $y$ axis clockwise through an angle $\phi$. 
    (d) Normalized magnetic susceptibility $\braket{\hat{J}_\varphi}/N$ (blue solid line)  and its normalized fluctuation $\Delta{}J_\varphi/\sqrt{N}$
    (magenta dashed line) on the $x-z$ plane, 
    given the atomic ensemble is in an one-axis spin squeezed state. Here $N=1000$.}
	\label{fig_quasiprob_magnetic_susceptibility}
\end{figure*}
It has attracted plenty attention in the endeavor to suppress the fluctuation of $\hat{J}_n$ 
and improve the single-atom 
resolution~\cite{zhang2012collective,hume2013accurate}. Others have worked to circumvent the single-atom resolution 
requirement. The main idea is to use nonlinear atomic interaction to suppress the fluctuation of 
$\hat{J}_n$, which is the so-called echo procedure. The echo procedure can utilize the same nonlinear interaction 
which is employed to generation multipartite entanglement on the first 
stage~\cite{linnemann2016quantumenhanced,hosten2016quantum,davis2016approaching,nolan2017optimal}. 
We find that measuring magnetic susceptibilities along two mutually orthogonal axes perpendicular to parameter field $\phi$ 
can produce parameter estimation precision beyond SQL. The importance of our work is two-fold. Firstly, it shows that neither single-atom 
resolution nor nonlinear interaction are necessary on final measurement stage of metrology to achieve precision improvement beyond SQL.
Secondly, it employs the most convenient measurement and thus is one step further to achieve metrology gain beyond SQL  promised 
by quantum mechanics in practice.

\section{One-axis spin squeezing}
We consider an atomic ensemble consist of $N$ atoms. They firstly are initialized to the coherent spin state along $x$ direction. 
We then use the one-axis spin squeezing Hamiltonian $\hat{J}_z^2$ to generate multipartite entanglement among the atoms. Following
the standard procedure of one-axis spin squeezing, a little adjustment along the $x$ direction would also be 
implemented~\cite{kitagawa1993squeezed,pezze2009entanglement},
\begin{equation}
	\ket{\psi_0} =e^{i\delta_\mathrm{adj}\hat{J}_x} e^{-it_\mathrm{s}\hat{J}_z^2}\ket{+}^{\otimes{}N}.
	\label{eq:oas}
\end{equation}
The adjustment is $\delta_\mathrm{adj}=\frac{1}{2}\arctan\frac{B}{A}$ with 
$A = 1-\left(\cos2t_\mathrm{s}\right)^{N-2}$ and $B = 4\sin{t_\mathrm{s}}\left(\cos{t_\mathrm{s}}\right)^{N-2}$. Here, $\ket{+}$ represents a single atomic spin
directing along the positive $x$ direction. We use the convention $\hbar=1$.
The squeezing time $t_\mathrm{s}$ (squeezing strength included) 
can be adjusted in experiment~\cite{linnemann2016quantumenhanced,hosten2016quantum,riedel2010atomchipbased,gross2010nonlinear,lucke2011twin,strobel2014fisher}.
That is the whole initialization stage. As shown in Fig.~\ref{fig_quasiprob_magnetic_susceptibility} (a) and (b), the effect of the spin squeezing is suppressing
fluctuation on the $z$ axis direction at the expense of increasing the fluctuation of spin along the two other directions. 

\section{Interference procedure}
The atomic spins would be sent through a magnetic field pointing to the positive $y$ direction. Let us assume the overall quantum channel is 
$\hat{U}_\phi = e^{-i\phi\hat{J}_y}$. The output state of the atomic ensemble is
\begin{equation}
	\ket{\psi_\phi} = e^{-i\phi\hat{J}_y}\ket{\psi_0}.
\end{equation}
By measuring atomic ensemble coming out of the magnetic field, we expect to estimate the value of $\phi$ and thus the value of the magnetic field.
\section{Magnetic susceptibility measurements along two directions.}
It is widely believed that there is always optimal measurement for a particular metrology scheme. In many cases, it can be proven that such kind of optimal 
measurement exists~\cite{braunstein1994statistical}. Others believe that adaptive feedback measurement can give the optimal information 
extraction~\cite{berry2000optimal,hentschel2010machineb,hentschel2011efficient,lovett2013differential,peng2020feedback}. Most of these measurement procedures 
are too complex to be realized on atomic ensembles. Magnetic susceptibility measurement is easy to carry out on atomic ensembles. In most cases, it is not optimal
measurement scheme. Our goal here is to use the most simple measurement to achieve as high as possible metrology precision. 

The reason that we choose two measurement directions of the magnetic susceptibility is of 
three fold. Firstly, the price to pay is small. We usually need to collect 
a sufficient amount data to infer $\phi$. Hence the initialization-interference-measurement procedure has to be repeated $R$ times with $R$ sufficiently big. We 
divide the $R$ experiments into two equal groups, of which each group measuring magnetic susceptibility along a different direction from the other group. Let us 
assume the worst case where magnetic susceptibility measurement on one of the direction provide no information of $\phi$. Then only $R/2$ data sample are of use.
The imprecision $\Delta\phi$ would increase by $\sqrt{2}$ which is  relatively small when $N$ of the order of $10^3\sim10^5$.
Secondly, there is actually a simple correspondence between the parameter field $\phi$ and 
the magnetic susceptibility on the $x-z$ plane of the Bloch sphere given the atom ensemble
is in an one-axis spin squeezed state \eqref{eq:oas},
c.f. Fig.~\ref{fig_quasiprob_magnetic_susceptibility}(d)
\begin{equation} 
	\braket{\hat{J}_\varphi}
	=\braket{\hat{J}_x}\sin\varphi+\braket{\hat{J}_z}\cos\varphi.
	\label{eq:magnetic_susceptibility_xz}
\end{equation} 
\begin{figure*}[!th]
	\includegraphics[width=\textwidth]{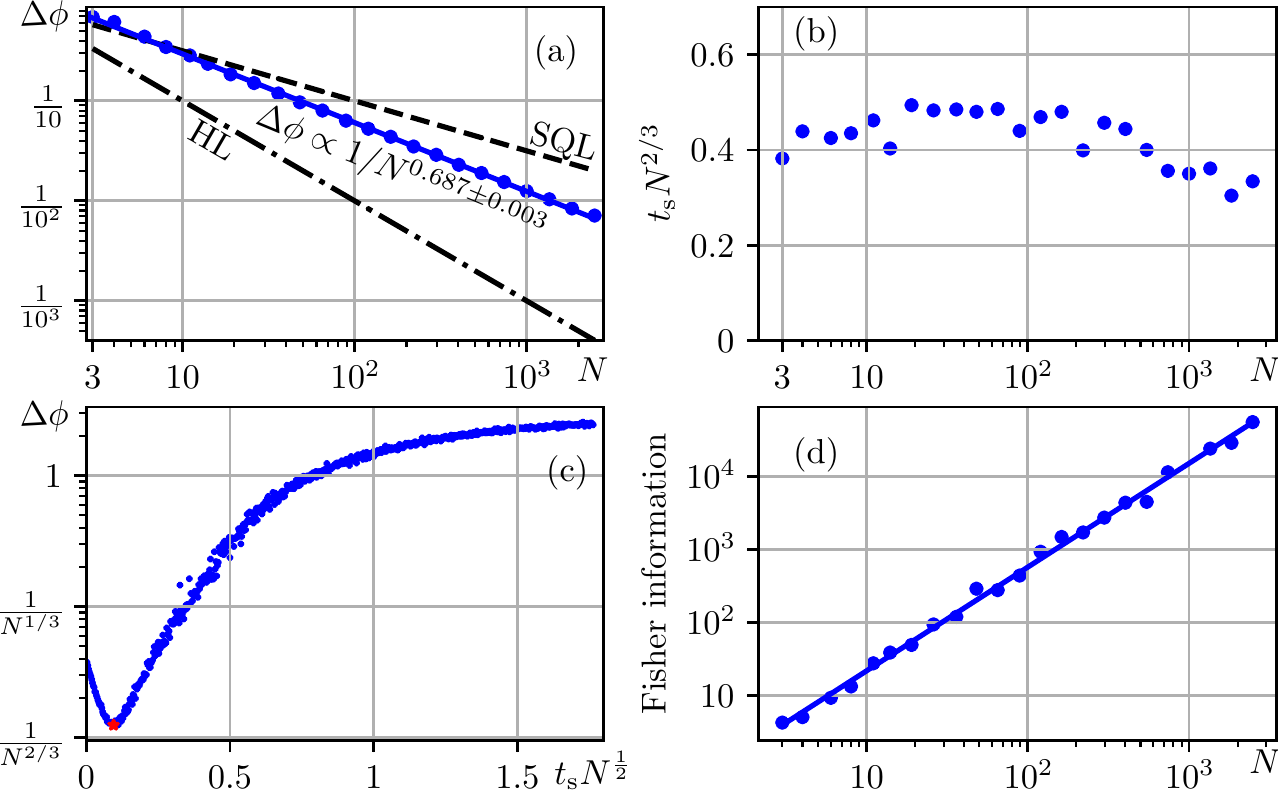}
	\caption{(Color online) Performance of the two-axis magnetic susceptibility measurement scheme for estimating $\phi$.
	(a) Scaling of the imprecision $\Delta\phi$ versus atomic ensemble size $N$.
	(b) The optimal squeezing time $t_\mathrm{s}$ for different atomic ensemble size $N$.
	(c) The imprecision $\Delta\phi$ given by 1000 atomic spins with different squeezing time. There is a single valley of 
	the $\Delta\phi$ versus $t_\mathrm{s}$ curve and the corresponding 
    optimal squeezing time (marked red star) is employed in our scheme.
	(d) Fisher information of one-axis spin squeezed state given optimized squeezing time as shown in (b).}
	\label{fig_performance}
\end{figure*} 

\noindent
There is only one single peak of the $\braket{\hat{J}_\varphi}$ curve which is located at 
$\varphi=\pi/2$. 
When the atomic ensemble initialized in an one-axis spin squeezed state comes
out of the $\hat{U}_\phi$ channel, the peak of $\braket{\hat{J}_\varphi}$ curve would shift
by $\phi$. $\braket{\hat{J}_\varphi}$ would be maximum if and only if $\varphi=\phi+\pi/2$
\begin{equation}
	\braket{\hat{J}_\varphi}
	=\sqrt{\braket{\hat{J}_x}^2+{\braket{\hat{J}_z}^2}}\sin(\varphi-\phi).
\end{equation} 
Hence $\phi$ can be calculated from $\braket{\hat{J}_x}$ and $\braket{\hat{J}_z}$, since
\begin{equation}
	\sin\phi
	=-\frac{\braket{\hat{J}_z}}{\sqrt{\braket{\hat{J}_x}^2+{\braket{\hat{J}_z}^2}}},
	\,\,\textrm{and}\,\,
	\cos\phi
	=\frac{\braket{\hat{J}_x}}{\sqrt{\braket{\hat{J}_x}^2+{\braket{\hat{J}_z}^2}}}.
	\label{eq_phi_jx_jz}
\end{equation}
In our scheme, we employ a slightly different evaluation formula from \eqref{eq_phi_jx_jz}
\begin{equation}
    \sin\phi_\mathrm{est}
    =-{j_z}/{\sqrt{j_z^2+j_x^2}}
	\,\,\textrm{and}\,\,
    \cos\phi_\mathrm{est}
    ={j_x}/{\sqrt{j_z^2+j_x^2}}.
    \label{eq_phi_jx_jz_singleShot}
\end{equation}
$j_z$ and $j_x$ are a pair of measurement results 
of magnetic susceptibility along the $x$ and $z$ direction respectively.
The two experiment measurement data produce one estimation 
$\phi_\mathrm{est}$ of the real parameter $\phi$. 

\subsection{Performance analysis}
We have simulated our metrology scheme for up to 2480 atomic spins. $R=2$ magnetic susceptibility
measurements are carried out for estimating each $\phi$, of which one is along the $x$ axis and the other along the $z$ axis. 
The two measurement results are submitted in \eqref{eq_phi_jx_jz_singleShot} to produce an estimation $\phi_\mathrm{est}$
of $\phi$. We choose the $M=1000$ parameter $\phi$ randomly  between 0 and $2\pi$. The performance of our scheme is estimated as 
\begin{equation}
	(\Delta\phi)^2 
	= \frac{R}{M}
	\sum_{i=1}^M
	\left[\mathrm{min}\left\{|\phi^{(i)}-\phi^{(i)}_\mathrm{est}|,2\pi-|\phi^{(i)}-\phi^{(i)}_\mathrm{est}|\right\}\right]^2.
\end{equation}
Note that $R$ in the above equation is to eliminate the metrology contribution of $R$ the size of data collected to estimate $\phi$.
$\phi$ is cyclic with period $2\pi$ and $\phi=0$ and $\phi=2\pi$ are considered to be of the same value.
We have optimized the squeezing time $t_\mathrm{s}$ to achieve the lowest $\Delta\phi$. 
Ref.~\cite{pezze2009entanglement} tells us that the entanglement behavior of one-axis spin squeezed state 
have a time scale of $1/N^{2/3}$
and its quantum Fisher information reach its maximum in the region of $t_\mathrm{s}\gtrsim1/N^{1/2}$. Hence 
we choose the range of $t_\mathrm{s}$ to be $[0,2/N^{1/2}])$
and optimize $t_\mathrm{s}N^{2/3}$ instead of $t_\mathrm{s}$.

As shown in Fig.~\ref{fig_performance}(a), our two-axis magnetic susceptibility measurement scheme can produce 
parameter estimation with imprecision below standard quantum limit for $N\gtrsim{}16$. The scaling $\Delta\phi$ with 
respect to atomic ensemble size is 
\begin{equation}
	\Delta\phi = \frac{1.43\pm0.02}{N^{0.687\pm0.003}}.
\end{equation}
The optimal $t_\mathrm{s}$ is around the value of $0.4/N^{2/3}$, as shown in  Fig.~\ref{fig_performance}(b). Given the 
optimized squeezing time $t_\mathrm{s}$, Fisher information $F$ of the corresponding one-axis spin squeezed state scales 
$F=0.833N^{1.42}$ with respect to $N$. It matches the scaling of $(\Delta{}\phi)^2$ with respect to $N$. 
\section{Summary and outlooks.}
The magnetic susceptibility fluctuations along two orthogonal axis cannot be both very small. This is prohibited by the Heisenberg
uncertainty relation. And sometimes $\Delta{}J_x$ as well as $\Delta{}J_z$ can be of the order of $\sqrt{N}$,
c.f. Fig.~\ref{fig_quasiprob_magnetic_susceptibility}(d). Our 
two-axis magnetic susceptibility measurement scheme can circumvent the worst effect of these fluctuation. 
By employing our scheme, neither 
sophisticated control schemes~\cite{berry2000optimal,hentschel2010machineb,hentschel2011efficient,lovett2013differential,peng2020feedback}
nor echo procedure induced by nonlinear interaction~\cite{linnemann2016quantumenhanced,hosten2016quantum,davis2016approaching,nolan2017optimal} is needed
to extract information from a quantum spin ensemble and SQL can be beaten. 
Besides, we need only two measurements to provide one estimation and thus small number of data is enough to ensure good estimation.
It is interesting to ask whether HL can be 
reached or approached with such a simple measurement scheme.
\begin{acknowledgments}
	This work was supported 
	National Natural Science Foundation of China (Grant Nos. T2121001, 11934018 and U1801661), 
	Strategic Priority Research Program of Chinese Academy of Sciences (Grant No. XDB28000000), 
	Key-Area Research and Development Program of GuangDong Province (Grant No. 2018B030326001),
	 Guangdong Provincial Key Laboratory (Grant No.2019B121203002),
the Science, Technology and Innovation Commission of Shenzhen Municipality (Grant No.KYTDPT20181011104202253), Grant No.2016ZT06D348.  
\end{acknowledgments}

\bibliography{Bib.bib}

\begin{thebibliography}{24}%
\makeatletter
\providecommand \@ifxundefined [1]{%
 \@ifx{#1\undefined}
}%
\providecommand \@ifnum [1]{%
 \ifnum #1\expandafter \@firstoftwo
 \else \expandafter \@secondoftwo
 \fi
}%
\providecommand \@ifx [1]{%
 \ifx #1\expandafter \@firstoftwo
 \else \expandafter \@secondoftwo
 \fi
}%
\providecommand \natexlab [1]{#1}%
\providecommand \enquote  [1]{``#1''}%
\providecommand \bibnamefont  [1]{#1}%
\providecommand \bibfnamefont [1]{#1}%
\providecommand \citenamefont [1]{#1}%
\providecommand \href@noop [0]{\@secondoftwo}%
\providecommand \href [0]{\begingroup \@sanitize@url \@href}%
\providecommand \@href[1]{\@@startlink{#1}\@@href}%
\providecommand \@@href[1]{\endgroup#1\@@endlink}%
\providecommand \@sanitize@url [0]{\catcode `\\12\catcode `\$12\catcode
  `\&12\catcode `\#12\catcode `\^12\catcode `\_12\catcode `\%12\relax}%
\providecommand \@@startlink[1]{}%
\providecommand \@@endlink[0]{}%
\providecommand \url  [0]{\begingroup\@sanitize@url \@url }%
\providecommand \@url [1]{\endgroup\@href {#1}{\urlprefix }}%
\providecommand \urlprefix  [0]{URL }%
\providecommand \Eprint [0]{\href }%
\providecommand \doibase [0]{http://dx.doi.org/}%
\providecommand \selectlanguage [0]{\@gobble}%
\providecommand \bibinfo  [0]{\@secondoftwo}%
\providecommand \bibfield  [0]{\@secondoftwo}%
\providecommand \translation [1]{[#1]}%
\providecommand \BibitemOpen [0]{}%
\providecommand \bibitemStop [0]{}%
\providecommand \bibitemNoStop [0]{.\EOS\space}%
\providecommand \EOS [0]{\spacefactor3000\relax}%
\providecommand \BibitemShut  [1]{\csname bibitem#1\endcsname}%
\let\auto@bib@innerbib\@empty
\bibitem [{\citenamefont {Giovannetti}\ \emph {et~al.}(2001)\citenamefont
  {Giovannetti}, \citenamefont {Lloyd},\ and\ \citenamefont
  {Maccone}}]{giovannetti2001quantumenhanceda}%
  \BibitemOpen
  \bibfield  {author} {\bibinfo {author} {\bibfnamefont {V.}~\bibnamefont
  {Giovannetti}}, \bibinfo {author} {\bibfnamefont {S.}~\bibnamefont {Lloyd}},
  \ and\ \bibinfo {author} {\bibfnamefont {L.}~\bibnamefont {Maccone}},\ }\href
  {\doibase 10.1038/35086525} {\bibfield  {journal} {\bibinfo  {journal}
  {Nature}\ }\textbf {\bibinfo {volume} {412}},\ \bibinfo {pages} {417}
  (\bibinfo {year} {2001})}\BibitemShut {NoStop}%
\bibitem [{\citenamefont {Jozsa}\ \emph {et~al.}(2000)\citenamefont {Jozsa},
  \citenamefont {Abrams}, \citenamefont {Dowling},\ and\ \citenamefont
  {Williams}}]{jozsa2000quantuma}%
  \BibitemOpen
  \bibfield  {author} {\bibinfo {author} {\bibfnamefont {R.}~\bibnamefont
  {Jozsa}}, \bibinfo {author} {\bibfnamefont {D.~S.}\ \bibnamefont {Abrams}},
  \bibinfo {author} {\bibfnamefont {J.~P.}\ \bibnamefont {Dowling}}, \ and\
  \bibinfo {author} {\bibfnamefont {C.~P.}\ \bibnamefont {Williams}},\ }\href
  {\doibase 10.1103/PhysRevLett.85.2010} {\bibfield  {journal} {\bibinfo
  {journal} {Phys. Rev. Lett.}\ }\textbf {\bibinfo {volume} {85}},\ \bibinfo
  {pages} {2010} (\bibinfo {year} {2000})}\BibitemShut {NoStop}%
\bibitem [{\citenamefont {Taylor}\ \emph {et~al.}(2013)\citenamefont {Taylor},
  \citenamefont {Janousek}, \citenamefont {Daria}, \citenamefont {Knittel},
  \citenamefont {Hage}, \citenamefont {Bachor},\ and\ \citenamefont
  {Bowen}}]{taylor2013biological}%
  \BibitemOpen
  \bibfield  {author} {\bibinfo {author} {\bibfnamefont {M.~A.}\ \bibnamefont
  {Taylor}}, \bibinfo {author} {\bibfnamefont {J.}~\bibnamefont {Janousek}},
  \bibinfo {author} {\bibfnamefont {V.}~\bibnamefont {Daria}}, \bibinfo
  {author} {\bibfnamefont {J.}~\bibnamefont {Knittel}}, \bibinfo {author}
  {\bibfnamefont {B.}~\bibnamefont {Hage}}, \bibinfo {author} {\bibfnamefont
  {H.-A.}\ \bibnamefont {Bachor}}, \ and\ \bibinfo {author} {\bibfnamefont
  {W.~P.}\ \bibnamefont {Bowen}},\ }\href {\doibase 10.1038/nphoton.2012.346}
  {\bibfield  {journal} {\bibinfo  {journal} {Nat. Photonics}\ }\textbf
  {\bibinfo {volume} {7}},\ \bibinfo {pages} {229} (\bibinfo {year}
  {2013})}\BibitemShut {NoStop}%
\bibitem [{\citenamefont {Schnabel}\ \emph {et~al.}(2010)\citenamefont
  {Schnabel}, \citenamefont {Mavalvala}, \citenamefont {McClelland},\ and\
  \citenamefont {Lam}}]{schnabel2010quantum}%
  \BibitemOpen
  \bibfield  {author} {\bibinfo {author} {\bibfnamefont {R.}~\bibnamefont
  {Schnabel}}, \bibinfo {author} {\bibfnamefont {N.}~\bibnamefont {Mavalvala}},
  \bibinfo {author} {\bibfnamefont {D.~E.}\ \bibnamefont {McClelland}}, \ and\
  \bibinfo {author} {\bibfnamefont {P.~K.}\ \bibnamefont {Lam}},\ }\href
  {\doibase 10.1038/ncomms1122} {\bibfield  {journal} {\bibinfo  {journal}
  {Nat. Commun.}\ }\textbf {\bibinfo {volume} {1}},\ \bibinfo {pages} {121}
  (\bibinfo {year} {2010})}\BibitemShut {NoStop}%
\bibitem [{\citenamefont {{The LIGO Scientific
  Collaboration}}(2013)}]{ligo2013enhanced}%
  \BibitemOpen
  \bibfield  {author} {\bibinfo {author} {\bibnamefont {{The LIGO Scientific
  Collaboration}}},\ }\href {\doibase 10.1038/nphoton.2013.177} {\bibfield
  {journal} {\bibinfo  {journal} {Nat. Photonics}\ }\textbf {\bibinfo {volume}
  {7}},\ \bibinfo {pages} {613} (\bibinfo {year} {2013})}\BibitemShut {NoStop}%
\bibitem [{\citenamefont {Braunstein}\ and\ \citenamefont
  {Caves}(1994)}]{braunstein1994statistical}%
  \BibitemOpen
  \bibfield  {author} {\bibinfo {author} {\bibfnamefont {S.~L.}\ \bibnamefont
  {Braunstein}}\ and\ \bibinfo {author} {\bibfnamefont {C.~M.}\ \bibnamefont
  {Caves}},\ }\href {\doibase 10.1103/PhysRevLett.72.3439} {\bibfield
  {journal} {\bibinfo  {journal} {Phys. Rev. Lett.}\ }\textbf {\bibinfo
  {volume} {72}},\ \bibinfo {pages} {3439} (\bibinfo {year}
  {1994})}\BibitemShut {NoStop}%
\bibitem [{\citenamefont {Giovannetti}\ \emph {et~al.}(2006)\citenamefont
  {Giovannetti}, \citenamefont {Lloyd},\ and\ \citenamefont
  {Maccone}}]{giovannetti2006quantum}%
  \BibitemOpen
  \bibfield  {author} {\bibinfo {author} {\bibfnamefont {V.}~\bibnamefont
  {Giovannetti}}, \bibinfo {author} {\bibfnamefont {S.}~\bibnamefont {Lloyd}},
  \ and\ \bibinfo {author} {\bibfnamefont {L.}~\bibnamefont {Maccone}},\ }\href
  {\doibase 10.1103/physrevlett.96.010401} {\bibfield  {journal} {\bibinfo
  {journal} {Phys. Rev. Lett.}\ }\textbf {\bibinfo {volume} {96}},\ \bibinfo
  {pages} {010401} (\bibinfo {year} {2006})}\BibitemShut {NoStop}%
\bibitem [{\citenamefont {Riedel}\ \emph {et~al.}(2010)\citenamefont {Riedel},
  \citenamefont {B{\"o}hi}, \citenamefont {Li}, \citenamefont {H{\"a}nsch},
  \citenamefont {Sinatra},\ and\ \citenamefont
  {Treutlein}}]{riedel2010atomchipbased}%
  \BibitemOpen
  \bibfield  {author} {\bibinfo {author} {\bibfnamefont {M.~F.}\ \bibnamefont
  {Riedel}}, \bibinfo {author} {\bibfnamefont {P.}~\bibnamefont {B{\"o}hi}},
  \bibinfo {author} {\bibfnamefont {Y.}~\bibnamefont {Li}}, \bibinfo {author}
  {\bibfnamefont {T.~W.}\ \bibnamefont {H{\"a}nsch}}, \bibinfo {author}
  {\bibfnamefont {A.}~\bibnamefont {Sinatra}}, \ and\ \bibinfo {author}
  {\bibfnamefont {P.}~\bibnamefont {Treutlein}},\ }\href {\doibase
  10.1038/nature08988} {\bibfield  {journal} {\bibinfo  {journal} {Nature}\
  }\textbf {\bibinfo {volume} {464}},\ \bibinfo {pages} {1170} (\bibinfo {year}
  {2010})}\BibitemShut {NoStop}%
\bibitem [{\citenamefont {Gross}\ \emph {et~al.}(2010)\citenamefont {Gross},
  \citenamefont {Zibold}, \citenamefont {Nicklas}, \citenamefont {Esteve},\
  and\ \citenamefont {Oberthaler}}]{gross2010nonlinear}%
  \BibitemOpen
  \bibfield  {author} {\bibinfo {author} {\bibfnamefont {C.}~\bibnamefont
  {Gross}}, \bibinfo {author} {\bibfnamefont {T.}~\bibnamefont {Zibold}},
  \bibinfo {author} {\bibfnamefont {E.}~\bibnamefont {Nicklas}}, \bibinfo
  {author} {\bibfnamefont {J.}~\bibnamefont {Esteve}}, \ and\ \bibinfo {author}
  {\bibfnamefont {M.~K.}\ \bibnamefont {Oberthaler}},\ }\href {\doibase
  10.1038/nature08919} {\bibfield  {journal} {\bibinfo  {journal} {Nature}\
  }\textbf {\bibinfo {volume} {464}},\ \bibinfo {pages} {1165} (\bibinfo {year}
  {2010})}\BibitemShut {NoStop}%
\bibitem [{\citenamefont {Lucke}\ \emph {et~al.}(2011)\citenamefont {Lucke},
  \citenamefont {Scherer}, \citenamefont {Kruse}, \citenamefont {Pezz\`e},
  \citenamefont {Deuretzbacher}, \citenamefont {Hyllus}, \citenamefont {Topic},
  \citenamefont {Peise}, \citenamefont {Ertmer}, \citenamefont {Arlt},
  \citenamefont {Santos}, \citenamefont {Smerzi},\ and\ \citenamefont
  {Klempt}}]{lucke2011twin}%
  \BibitemOpen
  \bibfield  {author} {\bibinfo {author} {\bibfnamefont {B.}~\bibnamefont
  {Lucke}}, \bibinfo {author} {\bibfnamefont {M.}~\bibnamefont {Scherer}},
  \bibinfo {author} {\bibfnamefont {J.}~\bibnamefont {Kruse}}, \bibinfo
  {author} {\bibfnamefont {L.}~\bibnamefont {Pezz\`e}}, \bibinfo {author}
  {\bibfnamefont {F.}~\bibnamefont {Deuretzbacher}}, \bibinfo {author}
  {\bibfnamefont {P.}~\bibnamefont {Hyllus}}, \bibinfo {author} {\bibfnamefont
  {O.}~\bibnamefont {Topic}}, \bibinfo {author} {\bibfnamefont
  {J.}~\bibnamefont {Peise}}, \bibinfo {author} {\bibfnamefont
  {W.}~\bibnamefont {Ertmer}}, \bibinfo {author} {\bibfnamefont
  {J.}~\bibnamefont {Arlt}}, \bibinfo {author} {\bibfnamefont {L.}~\bibnamefont
  {Santos}}, \bibinfo {author} {\bibfnamefont {A.}~\bibnamefont {Smerzi}}, \
  and\ \bibinfo {author} {\bibfnamefont {C.}~\bibnamefont {Klempt}},\ }\href
  {\doibase 10.1126/science.1208798} {\bibfield  {journal} {\bibinfo  {journal}
  {Science}\ }\textbf {\bibinfo {volume} {334}},\ \bibinfo {pages} {773}
  (\bibinfo {year} {2011})}\BibitemShut {NoStop}%
\bibitem [{\citenamefont {Strobel}\ \emph {et~al.}(2014)\citenamefont
  {Strobel}, \citenamefont {Muessel}, \citenamefont {Linnemann}, \citenamefont
  {Zibold}, \citenamefont {Hume}, \citenamefont {Pezz\`e}, \citenamefont
  {Smerzi},\ and\ \citenamefont {Oberthaler}}]{strobel2014fisher}%
  \BibitemOpen
  \bibfield  {author} {\bibinfo {author} {\bibfnamefont {H.}~\bibnamefont
  {Strobel}}, \bibinfo {author} {\bibfnamefont {W.}~\bibnamefont {Muessel}},
  \bibinfo {author} {\bibfnamefont {D.}~\bibnamefont {Linnemann}}, \bibinfo
  {author} {\bibfnamefont {T.}~\bibnamefont {Zibold}}, \bibinfo {author}
  {\bibfnamefont {D.~B.}\ \bibnamefont {Hume}}, \bibinfo {author}
  {\bibfnamefont {L.}~\bibnamefont {Pezz\`e}}, \bibinfo {author} {\bibfnamefont
  {A.}~\bibnamefont {Smerzi}}, \ and\ \bibinfo {author} {\bibfnamefont {M.~K.}\
  \bibnamefont {Oberthaler}},\ }\href {\doibase 10.1126/science.1250147}
  {\bibfield  {journal} {\bibinfo  {journal} {Science}\ }\textbf {\bibinfo
  {volume} {345}},\ \bibinfo {pages} {424} (\bibinfo {year}
  {2014})}\BibitemShut {NoStop}%
\bibitem [{\citenamefont {Berry}\ and\ \citenamefont
  {Wiseman}(2000)}]{berry2000optimal}%
  \BibitemOpen
  \bibfield  {author} {\bibinfo {author} {\bibfnamefont {D.~W.}\ \bibnamefont
  {Berry}}\ and\ \bibinfo {author} {\bibfnamefont {H.~M.}\ \bibnamefont
  {Wiseman}},\ }\href {\doibase 10.1103/physrevlett.85.5098} {\bibfield
  {journal} {\bibinfo  {journal} {Phys. Rev. Lett.}\ }\textbf {\bibinfo
  {volume} {85}},\ \bibinfo {pages} {5098} (\bibinfo {year}
  {2000})}\BibitemShut {NoStop}%
\bibitem [{\citenamefont {Hentschel}\ and\ \citenamefont
  {Sanders}(2010)}]{hentschel2010machineb}%
  \BibitemOpen
  \bibfield  {author} {\bibinfo {author} {\bibfnamefont {A.}~\bibnamefont
  {Hentschel}}\ and\ \bibinfo {author} {\bibfnamefont {B.~C.}\ \bibnamefont
  {Sanders}},\ }\href {\doibase 10.1103/PhysRevLett.104.063603} {\bibfield
  {journal} {\bibinfo  {journal} {Phys. Rev. Lett.}\ }\textbf {\bibinfo
  {volume} {104}},\ \bibinfo {pages} {063603} (\bibinfo {year}
  {2010})}\BibitemShut {NoStop}%
\bibitem [{\citenamefont {Hentschel}\ and\ \citenamefont
  {Sanders}(2011)}]{hentschel2011efficient}%
  \BibitemOpen
  \bibfield  {author} {\bibinfo {author} {\bibfnamefont {A.}~\bibnamefont
  {Hentschel}}\ and\ \bibinfo {author} {\bibfnamefont {B.~C.}\ \bibnamefont
  {Sanders}},\ }\href {\doibase 10.1103/physrevlett.107.233601} {\bibfield
  {journal} {\bibinfo  {journal} {Phys. Rev. Lett.}\ }\textbf {\bibinfo
  {volume} {107}},\ \bibinfo {pages} {233601} (\bibinfo {year}
  {2011})}\BibitemShut {NoStop}%
\bibitem [{\citenamefont {Lovett}\ \emph {et~al.}(2013)\citenamefont {Lovett},
  \citenamefont {Crosnier}, \citenamefont {{Perarnau-Llobet}},\ and\
  \citenamefont {Sanders}}]{lovett2013differential}%
  \BibitemOpen
  \bibfield  {author} {\bibinfo {author} {\bibfnamefont {N.~B.}\ \bibnamefont
  {Lovett}}, \bibinfo {author} {\bibfnamefont {C.}~\bibnamefont {Crosnier}},
  \bibinfo {author} {\bibfnamefont {M.}~\bibnamefont {{Perarnau-Llobet}}}, \
  and\ \bibinfo {author} {\bibfnamefont {B.~C.}\ \bibnamefont {Sanders}},\
  }\href {\doibase 10.1103/physrevlett.110.220501} {\bibfield  {journal}
  {\bibinfo  {journal} {Phys. Rev. Lett.}\ }\textbf {\bibinfo {volume} {110}},\
  \bibinfo {pages} {220501} (\bibinfo {year} {2013})}\BibitemShut {NoStop}%
\bibitem [{\citenamefont {Peng}\ and\ \citenamefont
  {Fan}(2020)}]{peng2020feedback}%
  \BibitemOpen
  \bibfield  {author} {\bibinfo {author} {\bibfnamefont {Y.}~\bibnamefont
  {Peng}}\ and\ \bibinfo {author} {\bibfnamefont {H.}~\bibnamefont {Fan}},\
  }\href {\doibase 10.1103/PhysRevA.101.022107} {\bibfield  {journal} {\bibinfo
   {journal} {Phys. Rev. A}\ }\textbf {\bibinfo {volume} {101}},\ \bibinfo
  {pages} {022107} (\bibinfo {year} {2020})}\BibitemShut {NoStop}%
\bibitem [{\citenamefont {Zhang}\ \emph {et~al.}(2012)\citenamefont {Zhang},
  \citenamefont {McConnell}, \citenamefont {{\'C}uk}, \citenamefont {Lin},
  \citenamefont {{Schleier-Smith}}, \citenamefont {Leroux},\ and\ \citenamefont
  {Vuleti{\'c}}}]{zhang2012collective}%
  \BibitemOpen
  \bibfield  {author} {\bibinfo {author} {\bibfnamefont {H.}~\bibnamefont
  {Zhang}}, \bibinfo {author} {\bibfnamefont {R.}~\bibnamefont {McConnell}},
  \bibinfo {author} {\bibfnamefont {S.}~\bibnamefont {{\'C}uk}}, \bibinfo
  {author} {\bibfnamefont {Q.}~\bibnamefont {Lin}}, \bibinfo {author}
  {\bibfnamefont {M.~H.}\ \bibnamefont {{Schleier-Smith}}}, \bibinfo {author}
  {\bibfnamefont {I.~D.}\ \bibnamefont {Leroux}}, \ and\ \bibinfo {author}
  {\bibfnamefont {V.}~\bibnamefont {Vuleti{\'c}}},\ }\href {\doibase
  10.1103/PhysRevLett.109.133603} {\bibfield  {journal} {\bibinfo  {journal}
  {Phys. Rev. Lett.}\ }\textbf {\bibinfo {volume} {109}},\ \bibinfo {pages}
  {133603} (\bibinfo {year} {2012})}\BibitemShut {NoStop}%
\bibitem [{\citenamefont {Hume}\ \emph {et~al.}(2013)\citenamefont {Hume},
  \citenamefont {Stroescu}, \citenamefont {Joos}, \citenamefont {Muessel},
  \citenamefont {Strobel},\ and\ \citenamefont
  {Oberthaler}}]{hume2013accurate}%
  \BibitemOpen
  \bibfield  {author} {\bibinfo {author} {\bibfnamefont {D.~B.}\ \bibnamefont
  {Hume}}, \bibinfo {author} {\bibfnamefont {I.}~\bibnamefont {Stroescu}},
  \bibinfo {author} {\bibfnamefont {M.}~\bibnamefont {Joos}}, \bibinfo {author}
  {\bibfnamefont {W.}~\bibnamefont {Muessel}}, \bibinfo {author} {\bibfnamefont
  {H.}~\bibnamefont {Strobel}}, \ and\ \bibinfo {author} {\bibfnamefont
  {M.~K.}\ \bibnamefont {Oberthaler}},\ }\href {\doibase
  10.1103/PhysRevLett.111.253001} {\bibfield  {journal} {\bibinfo  {journal}
  {Phys. Rev. Lett.}\ }\textbf {\bibinfo {volume} {111}},\ \bibinfo {pages}
  {253001} (\bibinfo {year} {2013})}\BibitemShut {NoStop}%
\bibitem [{\citenamefont {Linnemann}\ \emph {et~al.}(2016)\citenamefont
  {Linnemann}, \citenamefont {Strobel}, \citenamefont {Muessel}, \citenamefont
  {Schulz}, \citenamefont {{Lewis-Swan}}, \citenamefont {Kheruntsyan},\ and\
  \citenamefont {Oberthaler}}]{linnemann2016quantumenhanced}%
  \BibitemOpen
  \bibfield  {author} {\bibinfo {author} {\bibfnamefont {D.}~\bibnamefont
  {Linnemann}}, \bibinfo {author} {\bibfnamefont {H.}~\bibnamefont {Strobel}},
  \bibinfo {author} {\bibfnamefont {W.}~\bibnamefont {Muessel}}, \bibinfo
  {author} {\bibfnamefont {J.}~\bibnamefont {Schulz}}, \bibinfo {author}
  {\bibfnamefont {R.~J.}\ \bibnamefont {{Lewis-Swan}}}, \bibinfo {author}
  {\bibfnamefont {K.~V.}\ \bibnamefont {Kheruntsyan}}, \ and\ \bibinfo {author}
  {\bibfnamefont {M.~K.}\ \bibnamefont {Oberthaler}},\ }\href {\doibase
  10.1103/PhysRevLett.117.013001} {\bibfield  {journal} {\bibinfo  {journal}
  {Phys. Rev. Lett.}\ }\textbf {\bibinfo {volume} {117}},\ \bibinfo {pages}
  {013001} (\bibinfo {year} {2016})}\BibitemShut {NoStop}%
\bibitem [{\citenamefont {Hosten}\ \emph {et~al.}(2016)\citenamefont {Hosten},
  \citenamefont {Krishnakumar}, \citenamefont {Engelsen},\ and\ \citenamefont
  {Kasevich}}]{hosten2016quantum}%
  \BibitemOpen
  \bibfield  {author} {\bibinfo {author} {\bibfnamefont {O.}~\bibnamefont
  {Hosten}}, \bibinfo {author} {\bibfnamefont {R.}~\bibnamefont
  {Krishnakumar}}, \bibinfo {author} {\bibfnamefont {N.~J.}\ \bibnamefont
  {Engelsen}}, \ and\ \bibinfo {author} {\bibfnamefont {M.~A.}\ \bibnamefont
  {Kasevich}},\ }\href {\doibase 10.1126/science.aaf3397} {\bibfield  {journal}
  {\bibinfo  {journal} {Science}\ }\textbf {\bibinfo {volume} {352}},\ \bibinfo
  {pages} {1552} (\bibinfo {year} {2016})}\BibitemShut {NoStop}%
\bibitem [{\citenamefont {Davis}\ \emph {et~al.}(2016)\citenamefont {Davis},
  \citenamefont {Bentsen},\ and\ \citenamefont
  {{Schleier-Smith}}}]{davis2016approaching}%
  \BibitemOpen
  \bibfield  {author} {\bibinfo {author} {\bibfnamefont {E.}~\bibnamefont
  {Davis}}, \bibinfo {author} {\bibfnamefont {G.}~\bibnamefont {Bentsen}}, \
  and\ \bibinfo {author} {\bibfnamefont {M.}~\bibnamefont {{Schleier-Smith}}},\
  }\href {\doibase 10.1103/PhysRevLett.116.053601} {\bibfield  {journal}
  {\bibinfo  {journal} {Phys. Rev. Lett.}\ }\textbf {\bibinfo {volume} {116}},\
  \bibinfo {pages} {053601} (\bibinfo {year} {2016})}\BibitemShut {NoStop}%
\bibitem [{\citenamefont {Nolan}\ \emph {et~al.}(2017)\citenamefont {Nolan},
  \citenamefont {Szigeti},\ and\ \citenamefont {Haine}}]{nolan2017optimal}%
  \BibitemOpen
  \bibfield  {author} {\bibinfo {author} {\bibfnamefont {S.~P.}\ \bibnamefont
  {Nolan}}, \bibinfo {author} {\bibfnamefont {S.~S.}\ \bibnamefont {Szigeti}},
  \ and\ \bibinfo {author} {\bibfnamefont {S.~A.}\ \bibnamefont {Haine}},\
  }\href {\doibase 10.1103/PhysRevLett.119.193601} {\bibfield  {journal}
  {\bibinfo  {journal} {Phys. Rev. Lett.}\ }\textbf {\bibinfo {volume} {119}},\
  \bibinfo {pages} {193601} (\bibinfo {year} {2017})}\BibitemShut {NoStop}%
\bibitem [{\citenamefont {Kitagawa}\ and\ \citenamefont
  {Ueda}(1993)}]{kitagawa1993squeezed}%
  \BibitemOpen
  \bibfield  {author} {\bibinfo {author} {\bibfnamefont {M.}~\bibnamefont
  {Kitagawa}}\ and\ \bibinfo {author} {\bibfnamefont {M.}~\bibnamefont
  {Ueda}},\ }\href {\doibase 10.1103/physreva.47.5138} {\bibfield  {journal}
  {\bibinfo  {journal} {Phys. Rev. A}\ }\textbf {\bibinfo {volume} {47}},\
  \bibinfo {pages} {5138} (\bibinfo {year} {1993})}\BibitemShut {NoStop}%
\bibitem [{\citenamefont {Pezz{\'e}}\ and\ \citenamefont
  {Smerzi}(2009)}]{pezze2009entanglement}%
  \BibitemOpen
  \bibfield  {author} {\bibinfo {author} {\bibfnamefont {L.}~\bibnamefont
  {Pezz{\'e}}}\ and\ \bibinfo {author} {\bibfnamefont {A.}~\bibnamefont
  {Smerzi}},\ }\href {\doibase 10.1103/physrevlett.102.100401} {\bibfield
  {journal} {\bibinfo  {journal} {Phys. Rev.w Lett.}\ }\textbf {\bibinfo
  {volume} {102}},\ \bibinfo {pages} {100401} (\bibinfo {year}
  {2009})}\BibitemShut {NoStop}%
\end{thebibliography}%
\end{document}